\documentclass[11pt]{article}
\usepackage{a4wide}
\usepackage{latexsym}
\usepackage{epsfig}
\usepackage{amsmath}
\usepackage{amssymb}

\def\as{\alpha_s}

\newlength{\dinwidth} \newlength{\dinmargin}
\begin{document}

\thispagestyle{empty}

\begin{flushright}
  NIKHEF/2007-008\\
  ITP-UU-07/23\\
IMSc-2007/04/5
\end{flushright}

\vspace{1.5cm}

\begin{center}
 {\Large\bf Soft-collinear effects in prompt photon production}\\[1cm] 
  {\sc  Rahul Basu$^a$, Eric Laenen$^{b,c}$, Anuradha Misra$^d$, Patrick Motylinski$^b$\\
  \vspace{1.5cm}
  $^a$ {\it The Institute of Mathematical Sciences, CIT Campus,
  Taramani, Chennai 600 113, India}\\[0.4cm]
  $^b${\it NIKHEF Theory Group, Kruislaan 409, 1098 SJ Amsterdam, The Netherlands} \\[0.4cm]
  $^c${\it Institute for Theoretical Physics, Utrecht University\\
    Leuvenlaan 4, 3584 CE Utrecht, The Netherlands} \\[0.4cm]
  $^d$ {\it Department of Physics, University of Mumbai, Santacruz(E), Mumbai, 400\,098, India}\\
}
\end{center}

\vspace{2cm}


\begin{abstract}

\noindent{
We extend next-to-leading logarithmic threshold and joint 
resummation for prompt photon production
to include leading collinear effects.
The impact of these effects is assessed for both
fixed-target and collider kinematics. 
We find them in general to be small, but noticeable.}
\end{abstract}

\vspace*{\fill}

\newpage
\reversemarginpar

\section{Introduction}
\label{sec:introduction}

The perturbative QCD
description of many observables measured at colliders is plagued by
large corrections arising from soft and collinear parton emission,
even for fairly generic kinematical conditions.
For example, near threshold, large logarithmic corrections remain
\cite{Sterman:1987aj,Catani:1989ne}
after cancellation of singular virtual and real gluon contributions,
their large size being a result of the nearby threshold
restricting the real gluons to be soft. In terms of a (Mellin)
variable $N$, in terms of which threshold is approached by 
$N \rightarrow \infty$, such large threshold corrections 
take the form ($L = \ln N$),
\begin{equation}
  \label{eq:29}
  \alpha_s^i \sum_j^{2i}\, a_{ij} L^j\,,
\end{equation}
where the $a_{ij}$ depend in general on the process.
Another example 
\cite{Dokshitzer:1978yd,Parisi:1979se,Altarelli:1984pt,Collins:1981uk,Collins:1981va,Collins:1985kg}
 is when an identified part $F_P$ of a final state has 
acquired small transverse momentum by soft recoil 
($Q_T$) against the remaining, unmeasured part of the final state. Then the
perturbative expression for the differential cross section with
respect to $p_T$ of $F_P$ takes again the form of Eq.~\eqref{eq:29},
but with different coefficients $a_{ij}$ and with $L = \ln b$, $b$ being
the impact parameter  Fourier conjugate to $Q_T$.

Such large logarithmic corrections can be brought
under control by all-order resummation, and there exists a large literature 
demonstrating the viability, where applicable, of threshold, recoil as well 
as their joint resummation, for a wide variety of observables.
It is interesting to try to extend  all-order control
to classes of large terms beyond the logarithmic corrections.
One such new set consists of numerically 
large constants (``$\pi^2$ terms'') originating from the same
infrared-sensitive regions of those Feynman diagrams that also 
produce the logarithms \cite{Parisi:1980xd,Magnea:1990zb,Eynck:2003fn}. 

Another important class of potentially large terms, of soft-collinear origin, 
can be represented as
\begin{equation}
  \label{eq:30}
  \alpha_s^i \sum_j^{2i-1}\, d_{ij} \frac{\ln^jN}{N}\,.
\end{equation}
Their phenomenological importance was first demonstrated in
Ref.~\cite{Kramer:1996iq} in which the leading terms $j=2i-1$ 
were also summed to all orders for Higgs production
and Drell-Yan. The assessment of these terms was made
more meaningful in the context of a complete next-to-next-leading
order (NNLO)
\cite{Harlander:2002wh,Harlander:2001is,Anastasiou:2002yz,Anastasiou:2004xq,Ravindran:2003um,Ravindran:2003ia,Catani:2001ic} 
calculation, and a consistent next-to-next-leading logarithmic (NNLL) 
threshold-resummed  result \cite{Catani:2003zt}. It is not yet clear how to sum next-to-leading terms in \eqref{eq:30}.

In this paper we examine the impact of the leading terms in 
\eqref{eq:30} for a single particle inclusive observable, 
the $p_T$ spectrum of prompt photons produced in hadronic collisions.
We do this in the context of both a 
threshold \cite{Laenen:1998qw,Catani:1998tm,Catani:1999hs,Kidonakis:1999hq,Bolzoni:2005xn}
and joint \cite{Laenen:2000de,Laenen:2000ij,Sterman:2004yk,Li:1998is}
resummed calculation for this spectrum.

The paper is organized as follows. In section 2 we review briefly the
threshold and joint resummed prompt photon $p_T$ distribution. 
In section 3 we describe and motivate our extension to include
the leading $\alpha_s^k \tfrac{\ln^{2k-1}N}{N}$ terms. In section 4
we assess the numerical impact of these corrections, and 
we conclude in section 5.

\section{Threshold and joint resummation for prompt photon production}
\label{sec:thresh-joint-resumm}

We consider the inclusive $p_T$ distribution of prompt photons produced
in hadron-hadron collisions at center of mass (cm) energy $\sqrt{S}$
\begin{equation}
  \label{eq:el-proc}
  h_A(p_A) + h_B(p_B) \to \gamma(p_c)+X \>,
\end{equation}
where $h_{A,B}$ refer to the two incoming hadrons
 and $X$ to the unobserved part of the final state. 
The lowest order QCD processes producing the prompt photon
at partonic cm energy $\sqrt{s}$ are
\begin{equation}
  \label{eq:parton-proc}
  \begin{split}
    &q(p_a) + \bar q(p_b) \to \gamma(p_c) + g(p_d)\>, \\
    &g(p_a) +  q(p_b) \to \gamma(p_c) + q(p_d)\>.
  \end{split}
\end{equation}
The distance to threshold is customarily measured
by the variable $1-x_T^2$, where $x_T^2 = 4p_T^2/S$. At the parton level this
distance is given by $1-\hat{x}_T^2 = 1-4p_T^2/s$.
Below we review the result for 
the joint resummed prompt photon $p_T$ distribution.
At the end of this section we recall 
how the threshold resummed result may be derived 
from it.

The joint resummation formalism for prompt photon production
\cite{Laenen:2000de,Laenen:2000ij}
implements the notion that, in
the presence of soft QCD radiation
with summed transverse momentum ${\bf Q}_T$ of soft
recoiling partons, 
the actual transverse momentum produced
by the hard collision is not $\bf{p}_T$
but rather $\bf{p}\,'_T=\bf p_T-\bf Q_T/2$.
Stated more precisely: in the context of a refactorization analysis 
\cite{Laenen:2000ij} one can
identify a short-distance process at cm energy $Q$
that produces a prompt photon with momentum  $\bf{p}\,'_T$ in a recoiling frame. 
One defines accordingly $\tilde x^2_T = 4 {p^\prime}^2_T/Q^2$.
The extreme situation $Q_T = 2 p_T$ in which all transverse momentum is produced
through soft recoil leads to a singularity in the short-distance process, 
which we avoid by imposing an
upper limit $\bar{\mu}$ on $Q_T$ \cite{Laenen:2000de}. A recently proposed 
extension \cite{Sterman:2004yk} of joint resummation avoids this 
singularity.

The joint resummed $p_T$ distribution of prompt photons in hadronic
collisions is written as
\begin{eqnarray}
  \label{eq:6}
  {p_T^3 d \sigma^{({\rm resum})}_{AB\to \gamma+X} \over dp_T}
&=& \sum_{ij} \frac{p_T^4}{8 \pi S^2} \int_{\cal C} {dN \over 2 \pi i}\;
f_{i/A}(N,\mu_F) f_{j/B}(N,\mu_F)
\nonumber\\
&\ & \hspace{5mm} \times
\; \int_0^1 d\tilde x^2_T \left(\tilde x^2_T \right)^N\;
{|M_{ij}(\tilde x^2_T)|^2\over \sqrt{1-\tilde{x}_T^2}}\,
\;
C^{(ij\to \gamma k)}(\as(\mu),\tilde
x_T^2)
\nonumber \\
&& \hspace{10mm} \times
\int {d^2 {\bf Q}_T \over (2\pi)^2}\;
\Theta\left(\bar{\mu}-Q_T\right)
\left( \frac{S}{4 {\bf p}_T'{}^2} \right)^{N+1}\nonumber\\
&\ & \hspace{15mm} \times
\int d^2 {\bf b} \,
{\rm e}^{i {\bf b} \cdot {\bf Q}_T} \,
\exp\left[E_{ij\to \gamma k}\left( N,b,\frac{4 p_T^2}{\tilde
x^2_T},\mu_F \right)\right]\, .
\end{eqnarray}
Let us explain each of the terms on the right hand side of Eq.~\eqref{eq:6}.
The top line displays the moments of 
standard parton distribution functions, as well as the sum over
initial state parton flavors.
The next line contains the Mellin 
transform over the partonic scaling variable $\tilde{x}_T^2$ in the recoiling
frame, the Born amplitudes, 
and the $N$- and $b$-independent hard virtual corrections summarized in $C^{(ij\to \gamma k)}$.
The second to last line contains the integral over the recoil
momentum of the soft partons, 
as well as a kinematic factor linking
recoil and threshold effects.
The last line contains the Sudakov exponentials from 
initial  and final state partons, as well as soft wide-angle radiation
in combined Mellin-impact parameter space.

As indicated in the last line of Eq.~\eqref{eq:6}, large threshold and recoil 
logarithms, 
expressed through $\ln N$ and
$\ln b$, can be resummed into an exponential form. 
The perturbative exponential moment dependence at next-to-leading 
logarithmic (NLL) accuracy is given by
\begin{multline}
\label{eq:11}
E_{ij\rightarrow \gamma k}^{\rm PT} (N,b,Q,\mu,\mu_F) = \\
E^{\rm PT}_i (N,b,Q,\mu,\mu_F)  + E^{\rm PT}_j (N,b,Q,\mu,\mu_F) + F_k (N,Q,\mu) + G_{ijk} (N,\mu)\, .
\end{multline}
Let us discuss each of these terms in turn. The 
initial state perturbative exponent reads, in integral form
\begin{equation}
  \label{eq:39}
  E^{\rm PT}_i (N,b,Q,\mu,\mu_F)
= -\int_{Q^2/\chi^2}^{Q^2} {d k_T^2\over k_T^2}\; \left\{\ 
A_i\left(\as(k_T)\right)\; \ln\left(\frac{Q}{\bar{N}k_T} \right)\right\}
- 2\ln \bar{N}\int^{Q^2}_{\mu_F^2} {d k_T^2\over k_T^2}A_i\left(\as(k_T)\right)
\end{equation}
where
 $\mu,\mu_F$ are the renormalization and factorization scale, respectively.
The function $\chi(N,b)$ defines the $N$- and $b$-dependent scale of 
soft gluons to be included in the resummation,
and is chosen as \cite{Kulesza:2002rh}
\begin{equation}
  \label{eq:18}
\chi(N,b)=\bar{b} + \frac{\bar{N}}{1+\frac{\eta{\bar{b}}}{\bar{N}}}\; ,  
\end{equation}
where $\eta$ is a suitably chosen constant and
\begin{equation}
  \label{eq:17}
  \bar{N} = N e^{\gamma_E}, \quad
\bar{b} = bQ e^{\gamma_E}/2\,,
\end{equation}
with $\gamma_E$ the Euler constant. An older form used in \cite{Laenen:2000de}
\begin{equation}
  \label{eq:19}
  \chi(N,b)=\bar{b} + \bar{N}
\end{equation}
generates spurious subleading logarithms in $Q_T$ \cite{Kulesza:2002rh}.
We postpone elaborating on the integral in Eq.~\eqref{eq:39} to the next
section.
The final state jet exponent reads to NLL accuracy
\begin{equation}
  \label{eq:12}
  F_k (N,Q,\mu) \equiv \frac{1}{\alpha_s (\mu)} f^{(0)}_k (\lambda) +
f^{(1)}_k (\lambda,Q,\mu)\, ,
\end{equation}
where
\begin{equation}
\label{eq:25}
\lambda = b_0 \alpha_s(\mu^2)\ln \bar{N} \,.
\end{equation} 
The exponent associated with wide angle soft radiation is
\begin{equation}
  \label{eq:22}
G_{abc} (N) \equiv g^{(1)}_{abc} (\lambda)\, .  
\end{equation}
The functions $f^{(0,1)}_k$ and $g^{(1)}_{ijk} (\lambda)$  
as well as the functions $C^{(ij\to \gamma k)}$ \cite{Catani:1998tm,Catani:1999hs}
are listed in the Appendix.

A nonperturbative term must be added to the perturbative exponent in 
Eq.~\eqref{eq:11} 
in order to regularize the limit in which $Q_T$ is very small. As in Refs.~\cite{Laenen:2000de,Laenen:2000ij}
we take 
\begin{equation}
  \label{eq:37}
  E^{\mathrm{NP}}_{ij} = -\tfrac{1}{2} g_{\mathrm{NP}} b^2 \qquad
 ij = q\bar{q}, \, qg \,.
\end{equation}
The threshold-resummed result can now be derived by simply 
neglecting $\bf Q_T$ in the factor $( S/[4 |{\bf p}_T - {\bf Q}_T/2|^2])^{N+1}$
in Eq.~\eqref{eq:6}. Then the $\bf Q_T$ integral sets $\bf b$ to zero
everywhere, yielding the threshold-resummed result.

\section{Including leading $\ln N/N$ terms}
\label{sec:including-leading-ln}

The leading terms in Eq.~\eqref{eq:30} originate from both initial and final state
radiation, and to resum them we will use two different methods upon which we 
elaborate in this section.
There are moreover two classes of functions in momentum space at order
$\alpha_s^j$ that generate the leading
$\ln^{2j-1}N/N$ terms upon Mellin transformation. In terms of the
variable $z, \; 0<z<1$ which in the present case can represent either
$\hat{x}_T^2$ or $\tilde{x}_T^2$, 
one of the two classes is formed by 
the singular 
plus distributions $[\ln^{2j-1}(1-z)/(1-z)]_+$, the other by the singular
but integrable $\ln^{2j-1}(1-z)$. 
The $\ln N/N$ contributions from the former can be computed 
using the methods of \cite{Catani:1996yz} and can be found e.g. in 
Ref.~\cite{Kramer:1996iq}.
The $\ln N/N$ contributions from the latter can be generated at any order in 
perturbation theory by a simple
replacement in the resummed expression (see below in Eqs.~\eqref{eq:21}),
expanding the resulting expression to the desired order, and
keeping the leading term in Eq.~\eqref{eq:30}. Roughly speaking, 
the replacement 
is equivalent to exchanging at order $j$ one 
soft-collinear gluon (corresponding to one factor
$\alpha_s \ln^2 N$) for a hard-collinear one
(corresponding to a factor $\alpha_s \ln N/N$)
\begin{equation}
  \label{eq:31}
  \alpha_s^k \ln^{2k} N  \rightarrow   \alpha_s^k \frac{\ln^{2k-1}N}{N}\,.
\end{equation}
This replacement is in fact easily included in the existing threshold resummation
formulae. A preliminary study for prompt photon production was carried
out in Ref.~\cite{Mathews:2004pu}.
We employ this replacement method in fact for the final state related
$\alpha_s^k \ln^{2k-1}N/N$ terms.
It was pointed out in Refs.~\cite{Kulesza:2002rh,Kulesza:2003wn} that 
the initial state related $\alpha_s^k \ln^{2k-1}N/N$
terms could be generated in the context of joint resummation by extending
evolution of parton densities to a soft scale. We will use this
method as well, for the first time for a one-particle inclusive observable.
We now discuss the initial and final state $\ln N/N$ contributions
in turn.

\subsection{Initial state}
\label{sec:initial-state}

Our procedure for the initial state follows 
Refs.~\cite{Kulesza:2002rh,Kulesza:2003wn}, where the joint resummation for 
electroweak or Higgs boson production at mass $Q$ and transverse
momentum $Q_T$ was given. 
We recall  the key points here.
The integral form of the initial state NLL exponent \eqref{eq:39} can be written as
\begin{multline}
  \label{eq:13}
    E_i^{\rm PT}(N,b,Q,\mu,\mu_F)  =
-\int_{Q^2/\chi^2}^{Q^2} {d k_T^2\over k_T^2}\; \left\{\ 
A_i\left(\as(k_T)\right)\; \ln\left(\frac{Q}{k_T} \right)
+ B_i\left(\as(k_T)\right)\right\}\\
+ \int_{\mu_F^2}^{Q^2/\chi^2} {d k_T^2\over k_T^2}\; 
\left\{\ 
- \ln \bar N A_i\left(\as(k_T)\right) - B_i\left(\as(k_T)\right)
\right\}\,.
\end{multline}
The first term in this expression leads to 
\begin{equation}
  \label{eq:4}
      E_i^{\rm PT}(N,b,Q,\mu)  =
\frac{1}{\alpha_s (\mu)}h_i^{(0)} (\beta) +
h_i^{(1)} (\beta,Q,\mu)   \;  ,
\end{equation}
where
\begin{equation}
\label{eq:38}
\beta = b_0\, \alpha_s (\mu)
\ln \left( \chi \right) \, .
\end{equation}
We recall that the $\chi$ depends on $N$ and $b$ through Eq.~\eqref{eq:18}.
The functions $h_i^{(0,1)}$  are listed in the Appendix.

The second term represents flavor-conserving evolution to NLL accuracy
(the integrand consists of the $\ln N$ and constant terms 
for the anomalous dimension matrix $\gamma_{i/j}(N)$ for $j=i$)
from the hard scale $\mu_F$ to the soft scale $Q/\chi$. 
One now performs the replacement \cite{Kulesza:2002rh,Kulesza:2003wn} 
\begin{equation}
  \label{eq:15}
  - A_i(\as) \ln\left( \bar{N}\right) - B_i(\as) \; \longrightarrow \;
\gamma_{i/i}(N) (\as) \; ,
\end{equation}
that includes the leading, flavor-diagonal $\ln N/N$ effects
generated by the $k_T$ integral
(the $1/N$ part of $\gamma_{i/i}$ combines with the $\ln N$ terms).
In fact, one may go further and include the off-diagonal contributions via
the replacement
\begin{equation}
  \label{eq:16}
  \delta_{ig}\, \exp\left[ \frac{-A_g^{(1)}\ln\bar{N} -
B_g^{(1)}}{2\pi b_0} \, s(\beta) \right] \,  f_{g/H}(N ,\mu_F)\;
\longrightarrow \;
{\cal E}_{ik} \left(N,Q/\chi,\mu_F\right) \,
                 f_{k/H}(N ,\mu_F)\, .
\end{equation}
where $s(\beta) = \ln(1-2\beta)$ plus NLL corrections.
As a result, we can replace in Eq.~\eqref{eq:6} the combination
\begin{equation}
  \label{eq:10}
f_{i/A}(\mu_F,N) f_{j/B}(\mu_F,N) 
\exp\left[E^{\rm PT}_i (N,b,Q,\mu,\mu_F)  + E^{\rm PT}_j (N,b,Q,\mu,\mu_F)\right]
\end{equation}
by
\begin{equation}
\label{eq:2}
   {\mathcal C}_{i/A}(Q,b,N ) \;  {\mathcal C}_{j/B}(Q,b,N)\; 
\exp\left[E_i^{\rm PT}(N,b,\mu,Q)+E_j^{\rm PT}(N,b,\mu,Q)\right]
\end{equation}
where 
\begin{equation}
  \label{eq:5}
         {\mathcal C}_{i/H}(Q,b,N)
=   \sum_{k} {\cal E}_{ik} \left(N,Q/\chi,\mu_F\right) \,
                 f_{k/H}(N ,\mu_F) \; .
\end{equation}
The matrix $\mathcal{E}$ implements evolution from scale $\mu_F$ to scale $Q/\chi$,
and is normalized to be the unit matrix if these two scales are equal. 
Note that the dependence on $\mu_F$ cancels among the factors in Eq.~\eqref{eq:5}.

\subsection{Final state}
\label{sec:final-state}

Leading $\ln N/N$ effects arising from final state radiation
can be derived from the jet functions \cite{Sterman:1987aj,Kidonakis:1998bk}
that enter threshold or
joint resummed expressions for observables having final state
partons at lowest order.  
The integral form of the final state exponent $F_k$ in Eq.~\eqref{eq:11}  reads
\begin{equation}  
\label{eq:1}
   \int_0^1 dz \frac{z^{N-1}-1}{1-z} \left\{\int_{(1-z)^2Q^2}^{(1-z)Q^2} \frac{dq^2}{q^2} A_k(\alpha_s(q^2))
   +  B_k(\alpha_s((1-z)Q^2))\right\}\,.
\end{equation}
To include leading $\ln N/N$ dependence in $F_k(N,Q,\mu)$ we make the replacement
\cite{Kramer:1996iq,Catani:2003zt,Mathews:2004pu}
\begin{equation}
\label{eq:21}
\frac{z^{N-1} -1}{1-z} A^{(1)}_i  \rightarrow  \left[\frac{z^{N-1} -1}{1-z} - p_i z^{N-1}\right]\,
A^{(1)}_q + {\cal O}(\frac{1}{N^2})\,,
\end{equation}
where $p_q = 1 , p_g = 2$. The extra terms can be cast in a more
convenient form.  Using
\begin{equation}
\label{eq:35}
z^{N-1} = \frac{z^{N-1}-1 -(z^N-1)}{1-z}
\end{equation} 
and the replacement (accurate to NLL)
\begin{equation}
  \label{eq:40}
  z^{N-1}-1 \rightarrow -\theta\left(1-z-\frac{1}{\bar{N}}  \right)
\end{equation}
one finds
\begin{equation}
\label{eq:24}
F_k(N,Q,\mu) = \frac{1}{\alpha_s(\mu)} f^{(0)}_k(\lambda) + f^{(1)}_k(\lambda,Q,\mu) 
          +f'_k(\lambda,\alpha_s)+O(\alpha_s(\alpha_s\ln N)^n)\,,
\end{equation} 
where the extra terms $f\prime_k$ that include the leading $\ln N/N$ terms due to final state radiation
read
\begin{equation}
\label{eq:27}
f^\prime_q 
       =\frac{A^{(1)}_q}{ 2 \pi b_0}
  \exp\left(-\frac{\lambda}{\alpha_s b_0}\right)\left[\ln(1-2 \lambda)-\ln(1- \lambda)\right]\,,
\end{equation}
\begin{equation}
\label{eq:28}
f^\prime_g 
       =\frac{3 A^{(1)}_g}{ 2 \pi b_0}
  \exp\left(-\frac{\lambda}{\alpha_s b_0}\right)\left[\ln(1-2 \lambda)-\ln(1- \lambda)\right]\,.
\end{equation}
There is no leading $\ln N/N$ contribution arising from wide angle soft radiation.

As a result, we finally arrive at the following equation
for the joint resummed prompt photon hadroproduction 
$p_T$ spectrum in which leading soft-collinear effects are 
included: 
\begin{eqnarray}
  \label{eq:23}
    {p_T^3 d \sigma^{({\rm resum})}_{AB\to \gamma} \over dp_T}
&=& \frac{p_T^4}{8 \pi S^2}\ \sum_{ij}\  \int_{\cal C} {dN \over 2 \pi i}
\int d^2 {\bf b}\; {\rm e}^{i {\bf b} \cdot {\bf Q}_T} \;
\int {d^2 {\bf Q}_T \over (2\pi)^2}\;
\theta\left(\bar{\mu}-|{\bf Q}_T|\right)
\;\nonumber \\
&\ & \hspace{-25mm} \times\;
\int_0^1 d\tilde x^2_T\; \left(\tilde x^2_T \right)^N
{|M_{ij}(\tilde x^2_T)|^2\over \sqrt{1-\tilde{x}_T^2}}\;
C^{(ij\to \gamma k)}(\as(\mu),\tilde
x_T^2)\;\left( \frac{S}{4 |{\bf p}_T - {\bf Q}_T/2|^2} \right)^{N+1}
 \nonumber\\
&\ & \hspace{-20mm} \times
   {\mathcal C}_{i/A}(Q,b,N ) \;  {\mathcal C}_{j/B}(Q,b,N)\; 
\exp\left[E_i^{\rm PT}(N,b,\mu,Q)+E_j^{\rm PT}(N,b,\mu,Q)\right]\nonumber \\
&\ & \hspace{-15mm} \times
\exp\left[\frac{1}{\alpha_s(\mu)} f^{(0)}_k(\lambda) + f^{(1)}_k(\lambda,Q,\mu) 
          +f'_k(\lambda,\alpha_s)
+ g^{(1)}_{ijk} (\lambda)\right]\,.
\end{eqnarray}
As before, the corresponding threshold result may be obtained by neglecting
$-{\bf Q}_T/2$ in the last factor on the second line.

\section{Results}
\label{sec:results-1}

Here we study numerically the inclusion of the $\ln N/N$ terms for the
case of prompt photon production for two kinematic conditions:
those of $p\bar{p}$ collisions at the Tevatron at $\sqrt{S} = 1.96$ TeV
\cite{Abazov:2005wc,Acosta:2004bg},
and those of the $pN$ collisions in the 
E706 \cite{Apanasevich:2004dr} fixed target experiment with $E_{\mathrm{beam}} = 530$ GeV.
Our main aim is to assess the effect of such terms 
in relevant kinematic conditions, rather than provide optimized
and realistic theoretical calculations for comparison with
data (see Ref.~\cite{Aurenche:2006vj} for recent study). 
For instance, we do not include contributions from fragmentation processes,
which have recently been addressed in Ref.~\cite{deFlorian:2005wf}
and shown to be significant. Our assessments mainly consist
of comparing the same calculation with and without $\ln N/N$ terms. 

Our default choices for various input parameters are as follows. 
We use the GRV parton density set \cite{Gluck:1998xa}, corresponding to
$\alpha_s(M_Z)=0.114$, with the evolution code of
Ref.~\cite{Vogt:2004ns}, changing flavor number at $\mu = m_c\; (1.4
\,\mathrm{GeV})$ and $m_b \;(4.5 \, \mathrm{GeV})$.  We choose the
factorization and renormalization scale equal to $p_T$, and 
the non-perturbative parameter $g_{\mathrm{NP}}$ in
Eq.~\eqref{eq:37} equal to $1\,\mathrm{GeV}^2$. 
For the parameter $\chi$ we use 
the expression in Eq.~\eqref{eq:18}, following \cite{Kulesza:2002rh},
with $\eta = 1/4$ \footnote{Choosing $\eta =1 $ does not substantially modify results,
but choosing the form in Eq.~\eqref{eq:19}, which generates spurious subleading
recoil logs \cite{Kulesza:2002rh}, does lead to significant changes at larger $p_T$.
}.
For our joint-resummed results, we chose for Tevatron (E706) kinematics 
the cut-off $\bar{\mu}$ in Eq.~\eqref{eq:6} equal to $15\, (5)$ GeV. 
Regarding logarithmic accuracy, and unless otherwise stated we refer to LL when
using only $h^{(0)}_a$ in Eq.~\eqref{eq:4}, $f^{(0)}_k$ in Eq.~\eqref{eq:24},
and $\bar{C}^{(ij\to \gamma k)} = 1$ for the processes 
in \eqref{eq:parton-proc}; we refer to NLL 
when also including $h^{(1)}_a$ and $f^{(1)}_k$ and the
virtual corrections in \eqref{eq:34}. For the evolution
from scale $\mu_F$ to $Q/\chi$ in Eq.~\eqref{eq:16} we use the
full NLO anomalous dimension in all cases. 

\begin{figure}[htb!]
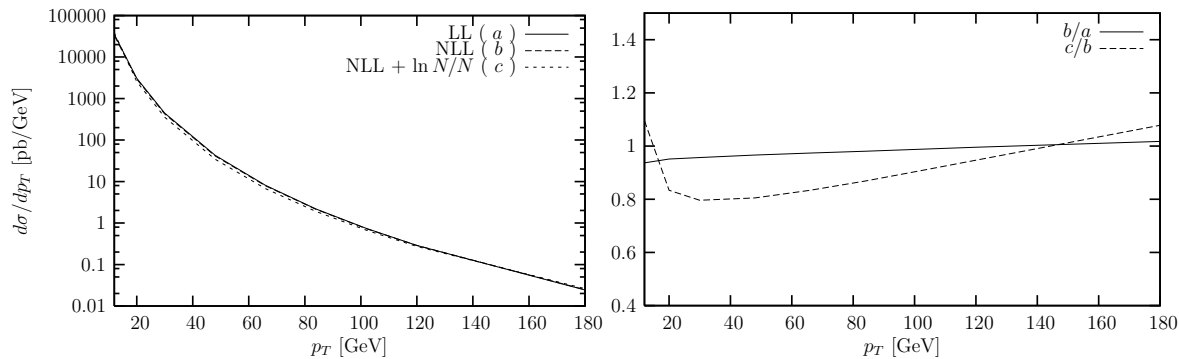

  \centering
  \includegraphics[width=0.49\textwidth]{FIG1a.epsi}
  \includegraphics[width=0.47\textwidth]{FIG1b.epsi}
 \caption{\small $\ln N/N$ contributions for Tevatron kinematics. 
Left pane: LL without $\ln N/N$ ($a$, solid),  NLL without $\ln N/N$ ($b$, dashed),
NLL with $\ln N/N$ ($c$, short-dashed). Right pane:
ratio of NLL to LL (solid), ratio of NLL with $\ln N/N$ to 
NLL without (dashed).} 
  \label{fig1}
\end{figure}

\begin{figure}[htb!]
  \centering
  \includegraphics[width=0.5\textwidth]{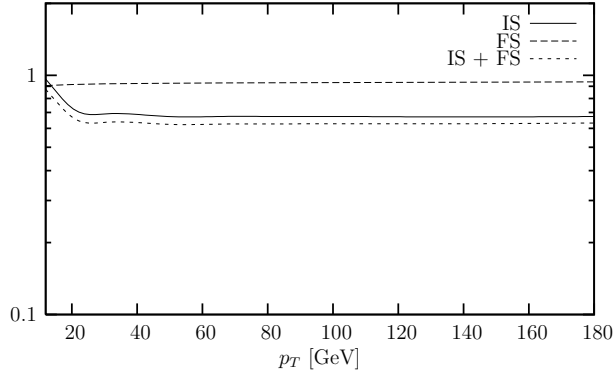}
  \caption{\small $\ln N/N$ effects for $q{\bar q}$ channel at LL, Tevatron kinematics.
Ratio to LL without $\ln N/N$ of initial state (solid) and final state (dashed) effects,
and both (short-dashed). }
  \label{fig2}
\end{figure}

\begin{figure}[htb!]
  \centering
  \includegraphics[width=0.5\textwidth]{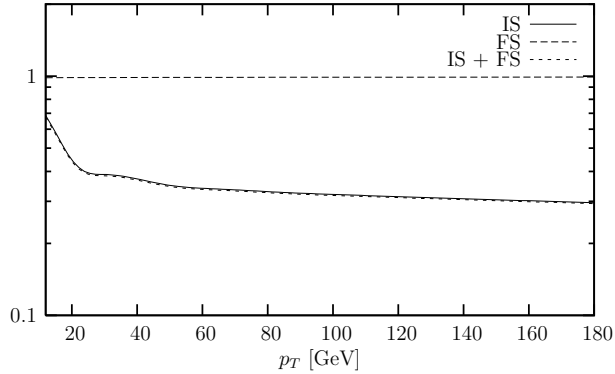}
  \caption{\small $\ln N/N$ effects for $qg$ channel at LL, Tevatron kinematics. Labels as
in Fig.~\ref{fig2}. }
  \label{fig3}
\end{figure}

Starting with Tevatron kinematics we compare in Figs.~\ref{fig1}-\ref{fig3}
results at LL and NLL accuracy, with and without the leading $\ln N/ N$ 
contribution for joint resummation.
For clarity we have here included the constant corrections in \eqref{eq:34} 
also for the LL case. Fig.~\ref{fig1} shows that the effect of the leading 
$\ln N/N$ is appreciable when compared to the effect of passing from LL to NLL, 
the latter difference being almost negligible. Inclusion of $\ln N/N$ effects  
leads to noticeable suppression for most of the $p_T$ range, and to enhancement
at very small and very large $p_T$.
To better understand the origin of these $\ln N/N$ suppressed
contributions, we examine in Figs.~\ref{fig2} and \ref{fig3} for each channel
in \eqref{eq:parton-proc} the contributions from the initial and final state.
We plot these contributions for the LL cross sections only to facilitate interpretation.
To help understand the results, we can 
expand the perturbative exponent in Eq.~\eqref{eq:11} to lowest order in
$\alpha_s$, keeping only the $\ln^2N$ and $\ln N/N$
terms
\begin{align}
  \label{eq:33}
& \mathrm{q\bar{q}}  :\quad:
\frac{\alpha_s}{\pi}\, \ln^2N\,\Big(2A_q^{(1)}-\tfrac{1}{2}A_g^{(1)}\Big)
+ \frac{\alpha_s}{\pi}\, \frac{\ln N}{N}\,\Big(2A_q^{(1)}-\tfrac{3}{2}A_g^{(1)}\Big) \\
& \mathrm{qg}  :\quad:
\frac{\alpha_s}{\pi}\, \ln^2N\,\Big(A_q^{(1)}+A_g^{(1)}-\tfrac{1}{2}A_q^{(1)}\Big)
+ \frac{\alpha_s}{\pi}\, \frac{\ln N}{N}\,\Big(A_q^{(1)}+3A_g^{(1)}-\tfrac{1}{2}A_q^{(1)}\Big) 
\end{align}
The expressions suggest that 
the initial state $\ln N/N$ terms enhance the cross section
for the $q\bar{q}$ and in particular the $qg$ channels, while
the final state $\ln N/N$ terms suppress it, again by an amount
that depends on the channel. The net result 
turns out to be suppression in the former channel and enhancement in the 
latter. 
These qualitative aspects are indeed borne out if we use the same method
to compute initial state $\ln N/N$ effects as we did 
for the final state in section \ref{sec:final-state} \footnote{The net
result in the $q\bar{q}$ is actually still enhancement, because the
contribution of the $f^\prime_{q,g}$ functions in ~\eqref{eq:27}, \eqref{eq:28} is 
very small.}.
In the present case however, the net $\ln N/N$ effect in both 
channels is suppression, indicating that the 
non-diagonal terms in the evolution matrix give a sizeable negative
contribution. 
Note that for Tevatron kinematics, when combining channels, 
the $qg$ channel dominates at low $p_T$, because the required momentum 
fractions are not too large. At large $p_T$, where parton
momentum fractions are larger, the valence-quark dominated $q{\bar q}$ 
channel takes over.

\begin{figure}[htb!]
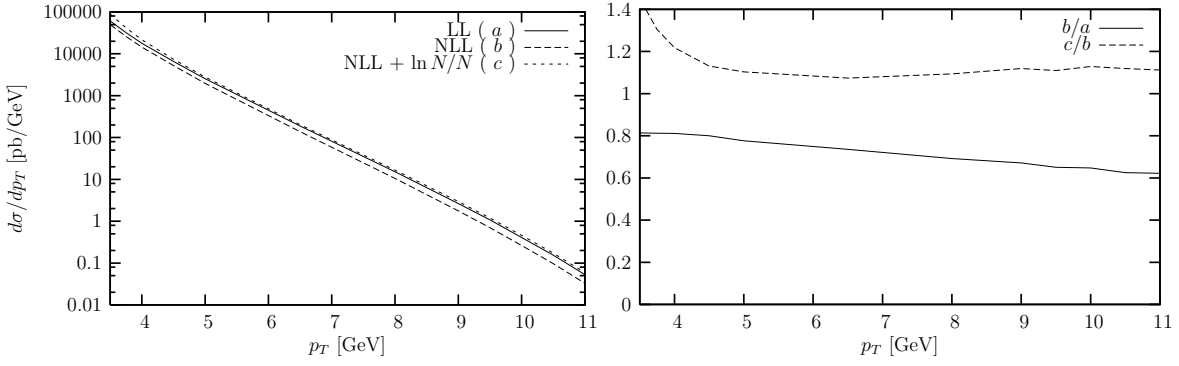

  \centering
  \includegraphics[width=0.49\textwidth]{FIG3a.epsi}
  \includegraphics[width=0.47\textwidth]{FIG3b.epsi}
  \caption{\small $\ln N/N$ contributions for E706 kinematics. Labels
 as in Fig.~\ref{fig1}.}
  \label{fig4}
\end{figure}

\begin{figure}[htb!]
  \centering
  \includegraphics[width=0.5\textwidth]{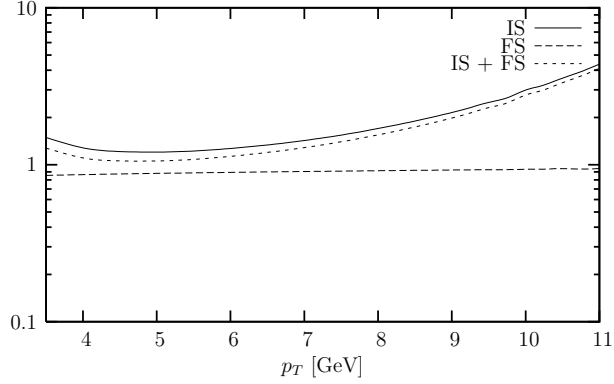}
  \caption{\small $\ln N/N$ effects for $q\bar{q}$ channel at LL, E706 kinematics.
Labels as in Fig.~\ref{fig2}. }
  \label{fig5}
\end{figure}

\begin{figure}[htb!]
  \centering
  \includegraphics[width=0.5\textwidth]{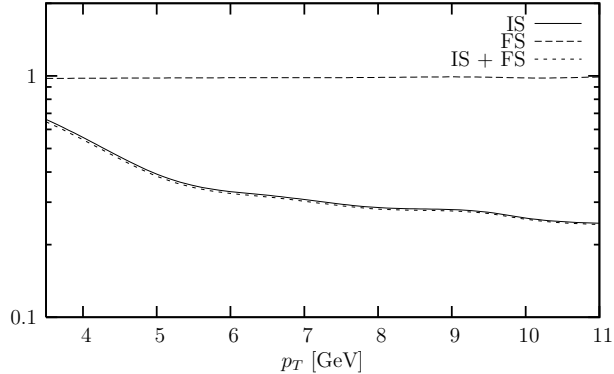}
  \caption{\small $\ln N/N$ effects for $qg$ channel at LL, E706 kinematics. 
Labels as in Fig.~\ref{fig2}. }
  \label{fig6}
\end{figure}

Turning to E706 kinematics we perform the same studies as we did
for the Tevatron. The results are shown in Figs.~\ref{fig4}-\ref{fig6}.
We observe an overall enhancement due to the $\ln N/N$ effects, somewhat
smaller than the change from LL to NLL. Both effects are more pronounced
than for the Tevatron. This is due both to a larger value of $\alpha_s$
as well as being closer to threshold in this fixed target kinematical regime.
Examining the effects per channel in Figs.~\ref{fig5} 
and \ref{fig6}, we now see a noticeable enhancement
from the initial state $\ln N/N$ effects in the $q\bar{q}$ channel, but still 
suppression in the $qg$ channel. Clearly the non-diagonal terms in the evolution
matrix play a significant role for the E706 case as well.

\begin{figure}[htb!]
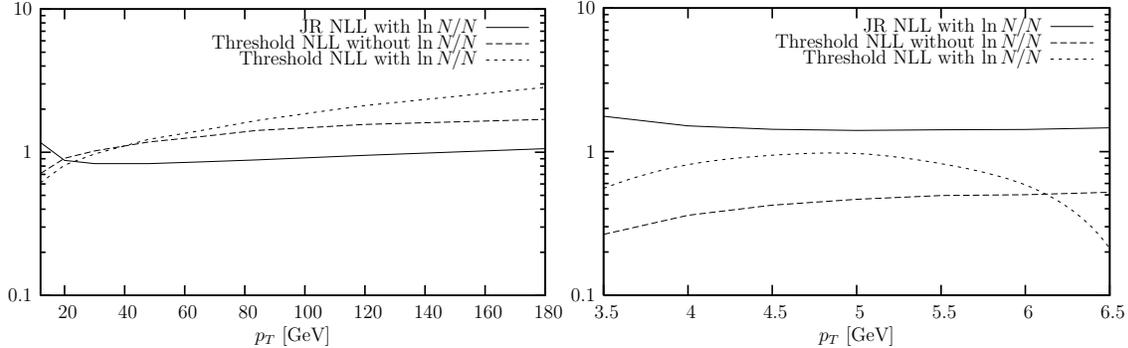

  \centering
  \includegraphics[width=0.46\textwidth]{FIG5a_ratio.epsi}
  \includegraphics[width=0.46\textwidth]{FIG5b_ratio.epsi}
  \caption{\small Comparison of joint resummation and threshold resummation effects,
ratios to NLL without $\ln N/N$ for Tevatron (left pane)
and E706 (right pane).}
\label{fig8}
\end{figure}

\begin{figure}[htb!]
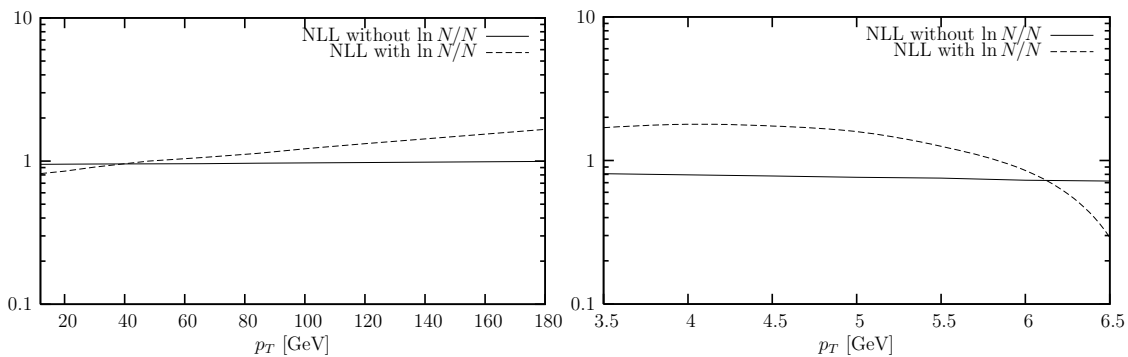

  \centering
  \includegraphics[width=0.46\textwidth]{FIG6a_ratio.epsi}
  \includegraphics[width=0.46\textwidth]{FIG6b_ratio.epsi}
  \caption{\small $\ln N/N$ effects in threshold resummation, 
for Tevatron (left pane) and E706 (right pane). Labels as in
Fig.~\ref{fig1} right pane.}
  \label{fig7}
\end{figure}

Next, we examine the differences between 
threshold and joint resummation. 
In Fig.~\ref{fig8} we compare resummed results directly
by showing the ratios with respect to the joint-resummed
$p_T$ distribution without $\ln N/N$ terms.
We see for Tevatron kinematics that the 
threshold resummed dominates the joint resummed at large $p_T$,
while at low $p_T$ the converse is true.
For the E706 case the threshold resummed results are
entirely below the joint-resummed ones.
The threshold resummed curves are shown separately in
Fig.~\ref{fig7}, which is analogous to the rightmost panels in 
Figs.~\ref{fig1} and \ref{fig4}.
For Tevatron kinematics the inclusion of $\ln N/N$ terms in
threshold resummation leads, as for joint resummation,
from suppression at small $p_T$ to enhancement at larger $p_T$,
but more noticeably.
For E706 kinematics, different from the joint resummation
case, the enhancement at small $p_T$ turns to suppression 
just below $p_T = 6$ GeV. The cross section even becomes
negative beyond $6.5$ GeV, which is due to the fact that
the nearness of the threshold drives the scale $Q/\chi$ in Eq.~\eqref{eq:16} 
effectively below the starting scale of the PDF evolution.

\section{Conclusions}
\label{sec:conclusions}

We have examined the effects of including terms of the form
\begin{equation}
\label{eq:26}
  \alpha_s^i \sum_j^{2i-1}\, d_{ij} \frac{\ln^jN}{N}\,.
\end{equation}
in joint-resummed and threshold-resummed prompt photon
$p_T$ distributions at both collider and fixed target kinematics,
at leading accuracy ($j=i$).
The complete structure of subleading terms of the form \eqref{eq:26} 
is still unknown. Note that we have not considered the fragmentation component 
of the prompt photon production cross section in our analysis \footnote{To do so would
require inclusion of more partonic subprocesses, each containing
a sum over color structures for the wide-angle soft radiation
component, as well as photon fragmentation functions~\cite{deFlorian:2005wf}.
Presumably, soft-collinear effects for the fragmentation component of prompt photon production could be included in a way analogous
to what we did in the present paper for the initial state: via adjustment of 
the resummed part, and evolution of the fragmentation functions.}.

To the extent that terms of the form \eqref{eq:26} arise from initial state radiation
effects, we used the method of Refs.~\cite{Kulesza:2002rh,Kulesza:2003wn}
to include them, now in a single-particle inclusive cross section.
Those arising from final state emission we included
by extending the jet function to leading $\ln N/N$
accuracy. Numerically we found the combined $\ln N/N$ terms to be 
comparable to NLL corrections, and dependent on kinematics either 
enhancing or suppressing.
The final state $\ln N/N$ contributions were particularly small,
while in the initial state the effects of non-leading $1/N$ effects
are appreciable, depending again on channel and kinematics. 
The flavour non-diagonal terms in the evolution matrix 
were found to be numerically significant, and the main source of
discrepancy with expectations based on simple approximations.
We conclude that, because the effects, though small, are non-negligible, 
understanding the structure of $\ln N/N$ terms better is a worthwhile pursuit.

\subsection*{Acknowledgments}

This work was supported by the Foundation for Fundamental Research of
Matter (FOM) and the National Organization for Scientific Research
(NWO). RB and AM would like to thank NIKHEF and EL and AM the IMSc in Chennai  
for local hospitality.

\appendix

\section{Appendix}

Here we list the exponents used in section \ref{sec:including-leading-ln}.
The initial state exponents \eqref{eq:4} involve
\begin{eqnarray}
\label{eq:14}
h_a^{(0)} (\beta) &=& \frac{A_a^{(1)}}{2\pi b_0^2}
\left[ 2 \beta + \ln(1-2 \beta) \right]\, ,\\
h_a^{(1)} (\beta,Q,\mu) &=&
\frac{A_a^{(1)} b_1}{2\pi b_0^3} \left[ \frac{1}{2} \ln^2 (1-2 \beta) +
\frac{2 \beta + \ln(1-2 \beta)}{1-2\beta} \right] + 
\frac{B_a^{(1)}}{\pi b_0}  \ln(1-2 \beta) \nonumber \\
&+& \frac{1}{2\pi b_0} \left[ A_a^{(1)}\ln \left( \frac{Q^2}{\mu^2} \right)
-\frac{A_a^{(2)}}{\pi b_0}\right] \;
\left[ \frac{2 \beta}{1-2\beta}+ \ln(1-2 \beta) \right] \; .
\end{eqnarray}
Here 
\begin{equation}
  \label{eq:9}
    A_a^{(1)} = C_a, \qquad
    A_a^{(2)} = 
\tfrac{1}{2}C_a \left[C_A\Bigg(\frac{67}{18}-\frac{\pi^2}{6}\Bigg)-\frac{10}{9}T_R N_F \right]
\end{equation}
where $C_q = C_F$ and $C_g = C_A$. Also
\begin{equation}
  \label{eq:20}
  B_q^{(1)} = -\frac{3}{4}C_F, \qquad
  B_g^{(1)} = -\pi b_0 \,.
\end{equation}
The final state exponents \eqref{eq:11} involve the functions
\begin{equation}
\label{eq:3}
f^{(1)}_a= -\frac{A^{(1)}_a}{2\pi b_0 \lambda}
          [(1-2 \lambda)\ln(1-2 \lambda) -2(1- \lambda)\ln(1- \lambda)]
\end{equation}   
\begin{multline}
\label{eq:7}
f^{(2)}_a = -\frac{A^{(1)}_a b_1}{2\pi b_0^3 }
            [\ln(1-2 \lambda) -2\ln(1- \lambda)
         +\frac{1}{2}\ln^2(1-2 \lambda)-\ln^2(1-\lambda)]\\
        +\frac{B^{(1)}_a}{ \pi b_0 }\ln(1-\lambda)
        -\frac{A^{(1)}_a\gamma_E}{\pi b_0 }[\ln(1-\lambda)-\ln(1-2\lambda)]
      -\frac{A^{(2)}_a}{2\pi^2 b_0^2}[2 \ln(1-\lambda) - \ln(1-2\lambda)]\\
+\frac{A^{(1)}_a}{2\pi b_0}[2\ln(1- \lambda)- \ln(1-2\lambda)]\ln\frac{Q^2}{\mu^2}
\end{multline}
The wide-angle soft radiation exponents \eqref{eq:22} are
\begin{equation}
  \label{eq:8}
g^{(1)}_{q\bar{q}g}(\lambda) = -\frac{C_A}{\pi b_0}\ln (1-2\lambda) \ln 2,\qquad
g^{(1)}_{qgq}(\lambda) = -\frac{C_F}{\pi b_0}\ln (1-2\lambda) \ln 2
\end{equation}
In these equations
\begin{eqnarray}
b_0 &=& \frac{11 C_A - 4 T_R N_F}{12 \pi}\;\;\;\; , \;\;\;\;\;
b_1 \;=\; \frac{17 C_A^2-10 C_A T_R N_F-6 C_F T_R N_F}{24 \pi^2}\; .
\end{eqnarray}
where $T_R = 1/2$.
These expressions are obtained by expanding the perturbative functions 
$A_a(\alpha_s)$, $B_d(\alpha_s)$ and $D_{ab\rightarrow d \gamma}$ in powers of 
$\alpha_s$ ,
\begin{equation}
\label{eq:32}
A_a(\alpha_s) = \frac{\alpha_s}{\pi}A^{(1)}_a + 
              \left(\frac{\alpha_s}{\pi}\right)^2 A^{(2)}_a +  O(\alpha^3_s)
\end{equation}
and so on. 

Finally, the explicit forms of $C^{(ij\to \gamma k)}$ \cite{Catani:1998tm,Catani:1999hs} are
\begin{align}
  \label{eq:34}
  & C^{q\bar{q}\rightarrow \gamma g} = 1 + 
\frac{\alpha_s}{\pi}\Bigg[ 
-\frac{1}{2}(2C_F - C_A)\ln 2 + \frac{1}{2}K - K_q + 2\zeta(2) (2C_F - \frac{1}{2}C_A)\\
& \hspace{0.5cm} + \frac{5}{4}(2C_F - \frac{1}{2}C_A)\ln^2 2 + \frac{3}{2}C_F (-\ln 2)
- \pi b_0\ln\frac{2p_T^2}{\mu^2} \Bigg] \\
  & C^{qg\rightarrow \gamma q} = 1 + 
\frac{\alpha_s}{\pi}\Bigg[ 
-\frac{1}{10}(C_F - 2C_A)\ln 2 - \frac{1}{2}K_q  + \frac{\zeta(2)}{10} (2C_F + 19C_A) \\
& \hspace{0.5cm} + \frac{1}{2}C_F\ln^2 2 + \frac{3}{4}(C_F+\pi b_0) (-\ln 2)
- \pi b_0\ln\frac{2p_T^2}{\mu^2} \Bigg] 
\end{align}
where
\begin{align}
  \label{eq:36}
  & K = C_A \left(\frac{67}{18}-\frac{\pi^2}{6} \right) - \frac{10}{9}T_R N_F
  & K_q = \left(\frac{7}{2}-\frac{\pi^2}{6} \right)C_F
\end{align}
We note that there is no factorization scale dependence in $h_a^{(1)}$ and the coefficient
functions in Eq.~\eqref{eq:34} because of complete evolution from scale $\mu_F$ to $Q/\chi$
in Eqs.~\eqref{eq:10},\eqref{eq:2} and \eqref{eq:5}.




\end{document}